\documentclass[prd,aps,12pt,nofootinbib,superscriptaddress]{revtex4-1}

\usepackage{amsmath,amssymb}
\usepackage[normalem]{ulem}

\usepackage{graphicx}

\usepackage{subfigure}

\usepackage{physics}

\usepackage{xcolor}

\newcommand{\beq}{\begin{equation}}
\newcommand{\eeq}{\end{equation}}

\newcommand{\ber}{\begin{eqnarray}}
\newcommand{\eer}{\end{eqnarray}}

\def\ie{\rm i.e.~}

\def\0{{\scriptscriptstyle (0)}}
\def\1{{\scriptscriptstyle (1)}}
\def\q{{x}}
\def\l{\left}
\def\r{\right}
\def\d{{\scriptscriptstyle E}}

\def\g{{g}}

\def\oml{\Omega_{\ell}}
\def\om0{\Omega_{m0}}
\def\Om{\Omega_{m}}
\def\Os{\Omega_{\sigma}}

\begin{document}

 \title{Versatile parametrization of the perturbation growth rate on the phantom brane}

 \author{Alexander Viznyuk}\email{viznyuk@bitp.kiev.ua}
 \affiliation{Bogolyubov Institute for Theoretical Physics,  Kiev 03143, Ukraine} %
 \affiliation{Department of Physics, Taras Shevchenko National University, Kiev 03022, Ukraine}
 \affiliation{Department of Theoretical and Mathematical Physics, Kiev Academic University, Ukraine} %

 \author{Satadru Bag}\email{satadru@iucaa.in}
 \affiliation{Inter-University Centre for Astronomy and Astrophysics, Pune 411007, India} %

 \author{Yuri Shtanov}\email{shtanov@bitp.kiev.ua}
 \affiliation{Bogolyubov Institute for Theoretical Physics, Kiev 03143, Ukraine} %
 \affiliation{Department of Physics, Taras Shevchenko National University, Kiev 03022, Ukraine}

 \author{Varun Sahni}\email{varun@iucaa.in}
 \affiliation{Inter-University Centre for Astronomy and Astrophysics, Pune 411007, India}

 \begin{abstract}
We derive an analytical expression for the growth rate of matter density perturbations on the
phantom brane (which is the normal branch of the Dvali--Gabadadze--Porrati model). This model is characterized by a
phantomlike  effective equation of state for dark energy at the present epoch.
It agrees very well with observations. We
demonstrate that the traditional parametrization $f=\Omega_m^\gamma$ with a quasiconstant growth index $\gamma$ is not successful in this case. Based on a power series expansion at large redshifts,
we propose a different parametrization for this model: $f=\Omega_m^\gamma\left(1+\frac{b}{\ell H}\right)^\beta$, where $\beta$ and $b$ are constants. Our numerical simulations demonstrate that this new parametrization describes the growth rate with great accuracy---the maximum error being $\leq 0.1\%$  for parameter values consistent with observations.

 \end{abstract}

\maketitle

\tableofcontents


\section{Introduction}\label{sec: Intro}

According to the braneworld paradigm (see \cite{Maartens:2010ar, Novosyadlyj:2015zpa} for  reviews), our universe is a four-dimensional hypersurface (the ``brane'') embedded in a
five-dimensional spacetime (the ``bulk''). In this scenario, the matter and gauge fields of the standard model are confined  to the brane, while gravity can propagate in the extra dimension.

An important class of braneworld models, known as the Dvali--Gabadadze--Porrati (DGP) model, contains the so-called `induced-gravity' term in
the action for the brane, which modifies gravity on relatively large spatial scales \cite{Collins:2000yb, DGP, Shtanov:2000vr}.  Depending upon the embedding of the brane in the
bulk space, this model has two branches of cosmological solutions: the ``self-accelerating'' branch and
the ``normal'' branch \cite{Deffayet:2000uy}. The self-accelerating branch can describe cosmology with late-time
acceleration without bulk and brane cosmological constants \cite{Deffayet:2001pu}, but it is plagued by the existence of ghost excitations \cite{Ghosts}.
On the normal branch, late-time
acceleration can be realized via the brane tension, which plays the role of a cosmological constant on the brane. No ghosts appear in this case.

Because of the ghost problem, the self-accelerating branch is of limited interest, while the normal
branch is physically viable and consistent with current cosmological observations.
Describing the cosmological solution of the normal branch in terms of effective dark energy, one notes that it has a phantomlike effective equation of state $w_{\rm eff} < -1$ at late times, but does not run
into a big-rip future singularity \cite{Sahni:2002dx, Alam_Sahni, Lue:2004za,Lazkoz:2006gp,Alam:2016wpf}. In view of this property, the term ``phantom brane'' was proposed for the normal branch in \cite{Bag:2016tvc}.
This model will be in the focus of the present investigation.

In the literature, the phantom braneworld model has been confronted against various distance  measures \cite{Lazkoz:2006gp, Alam:2016wpf}, specifically, from Type~Ia supernovae, baryon acoustic oscillations (BAO) and CMB observations. A recent study \cite{Alam:2016wpf} showed that these distance measures
are consistent with the presence of an extra dimension, and constrain the brane parameter,
defined in \eqref{Omega_m q brane definition}, to be $\oml \lesssim 0.1$ at $1\sigma$; also see
\cite{Alam_Sahni:2006,Lombriser:2009xg,Xu:2013ega,Bhattacharya:2018}.
An important feature of the phantom brane is that its expansion rate
{\em is slower than} $\Lambda$CDM, \ie $H(z)\vert_{\rm brane} < H(z)\vert_{\Lambda{\rm CDM}}$.
This intriguing property allows the braneworld to better account for measurements of
$H(z)$ at $z \sim 2$, reported in \cite{BAO}, which appear to be in some tension with
$\Lambda$CDM; also see \cite{sss14,shs18}.

To test braneworld cosmology at the linear perturbative level, one needs to know
the behaviour of matter density perturbations in this model.
This is usually described in terms of the growth rate
 $f = \dd \ln \delta_m/\dd \ln a$, where
$\delta_m=\delta\rho_m/\rho_m$ is the matter density contrast. For the $\Lambda$CDM model and for a large variety of dynamical dark energy models with slowly varying $w$, the growth rate can be approximated as
 \beq\label{growth rate GR}
f=\Omega_{m}^{\gamma} \quad \text{with} \quad \Omega_m = \frac{8 \pi G \rho_m}{3 H^2} \,,
\eeq
where $\gamma$ is the growth index \cite{Peebles:1980, Wang:1998gt, Polarski:2016ieb}. For low redshifts, $\gamma$ is a slowly varying function of $z$, close to some constant $\gamma_0$. The value of $\gamma_0$ depends on the equation of state $w$, and thus the growth index can be used to discriminate between different models of the dynamical dark energy. For example,  when $w=-1$ (as in the $\Lambda$CDM model), we have $\gamma_0=6/11$.

The parametrization \eqref{growth rate GR} can also be applied to the description of perturbations for some modified gravity theories. In particular, the behavior of the growth index in the self-accelerating branch of the  DGP braneworld model is given by \eqref{growth rate GR} with $\gamma_0=11/16$ \cite{Linder:2007hg, Wei:2008ig, Gong:2008fh, Phong:2012ig}. This approximation can be improved  assuming that the growth index is a function of $z$. Successful parametrization of the growth index allows one to reduce the discrepancy between \eqref{growth rate GR} and the numerical solution for the growth rate to a relative value below $0.04\%$, as reported by \cite{Ishak:2009qs}, and even below $0.028\%$ \cite{Chen:2009ak}.

It was argued that parametrization in terms of
the growth index $\gamma$ alone is not enough to get a satisfactory growth rate for  some modified models of gravity \cite{Resco:2017jky}. Discussion about a possible universal parametrization for modified gravity models continues, and, in this paper, we hope to provide  additional inputs to this debate.

In contrast to \cite{Resco:2017jky}, we restrict our investigation to a specific model of modified
gravity, the phantom brane.
Our aim will be to find an analytic expression for the growth rate on the phantom brane and to study
whether the parametrization \eqref{growth rate GR} works in this case.\footnote{ To the best of our knowledge, this will be the first attempt to find an analytical expression for the growth rate on the normal branch of the braneworld model.}  We will find that a parametrization involving
 the growth index fails to describe the exact (numerical) solution for the growth rate in this model.  This is due to the fact that the quantity $\Omega_m$ in our braneworld is a nonmonotonic function of
redshift. Therefore,
the exact growth rate on the phantom brane also becomes a
multivalued function of $\Omega_m$ for large values of $z \geq z_p$ [$z_p = {\cal O} (1)$].
This behaviour cannot be accommodated by
 $f = \Omega_m^\gamma$ which is a single-valued function of $\Omega_m$.

Our paper is organized as follows. In Sec.~\ref{sec: brane background}, we discuss the peculiarities of the background cosmological evolution on the phantom brane. In Sec.~\ref{sec: brane growth index series}, we perform a series expansion of the growth rate in this model in the asymptotic past, thus getting an approximate solution valid for $z\gg 1$. In Sec.~\ref{sec: parametr gamma beta}, we describe the process of finding a parametrization that fits the asymptotic expansion in the asymptotic past, and compare the resulting parametrization with the numerical solution for the growth rate. In Sec.~\ref{sec:future}, we study solutions of the growth function in the asymptotic future.  A successful ansatz, valid for $z\geqslant 0$, is described in Sec.~\ref{sec:numerical}. Our results are summarized in Sec.~\ref{sec: conclusion}.


\section{Background cosmological evolution}\label{sec: brane background}

The phantom brane is the normal branch of the braneworld cosmological solution. For a spatially flat brane embedded in the flat bulk space-time, the following expression describes the evolution of the Hubble parameter $H=\dot{a}/a$ in this model
\cite{Bag:2016tvc}:
\beq \label{background normal branch}
H= \sqrt{\frac{\rho_m+\sigma}{3m^2}+\frac{1}{\ell^2}}-\frac{1}{\ell} \, .
\eeq
Here, $\rho_m$ is the energy density of matter,\footnote{In the following, we are interested the evolution of cosmological perturbations commencing in the matter-dominated epoch, when contributions from
radiative degrees of freedom can be neglected.} $1/m^2=8\pi G$ is the gravitational constant, $\sigma$ is the brane tension and $\ell$ is the length scale which describes the interplay between the bulk and brane gravity.

Equation \eqref{background normal branch} can be written in a form that expresses $\rho_m$ in terms of $H$:
\beq \label{background normal branch ell H}
\frac{\rho_m + \sigma }{ 3 m^2} = H^2 \l(1+\frac{2}{\ell H}\r)
\, .
\eeq
One can see from \eqref{background normal branch} and \eqref{background
normal branch ell H} that  cosmological evolution on the phantom brane
has the general-relativistic limit
when $\ell\rightarrow \infty$. In this case, the braneworld model is
equivalent to the $\Lambda$CDM model with $\sigma/m^2$ playing the role
of $\Lambda$-term. We
are interested in exploring effects stemming from large, but
not infinite, values of $\ell$.

It is convenient for further purposes to introduce the dimensionless variables
\beq \label{Omega_m q brane definition}
\Omega_{m,0} =\frac{\rho_{m,0}}{3m^2 H_0^2}\,,\qquad\Omega_\ell =\frac{1}{\ell^2 H_0^2}\,,\qquad\Omega_\sigma
=\frac{\sigma}{3m^2 H_0^2}\,,
\eeq
where $H_0$ and $\rho_{m,0}$ are the values of
Hubble parameter and matter density at the present epoch.
In terms of these variables, the evolution of the Hubble parameter becomes
\beq \label{background normal branch dimensionless}
h \equiv \frac{H}{H_0}
= \sqrt{\Omega_{m,0}(1+z)^3 + \Omega_\sigma + \Omega_\ell} - \sqrt{\Omega_\ell}\,.
\eeq
Here, $z$ is the cosmological redshift, related to the cosmological scale factor as $1+z=a_0/a$.
At the present epoch ($z=0$), $H=H_0$, which provides the constraint
\beq \label{Omega constraint}
\Omega_\sigma
= 1 - \Omega_{m,0} + 2\sqrt{\Omega_\ell} \,.
\eeq

The quantity $\Omega_\ell$ parameterizes
deviations from the $\Lambda$CDM model. The general-relativistic limit
$\ell\rightarrow \infty$ is equivalent to $\Omega_\ell \rightarrow 0$.


\subsection{Properties of the effective dark energy}\label{sec: brane effective}

The background cosmological evolution in this braneworld model can well be described in terms of the effective dark energy. The energy density $\rho_\d$ and pressure $p_\d$ of the effective dark energy are defined via the following relations \cite{ss06}
\beq\label{brane effective}
H^2=\frac{\rho_m + \rho_\d}{3 m^2} \,,\qquad\dot{H}= -\,
\frac{1}{2m^2}\left(\rho_m+ \rho_\d + p_\d \right) \,.
\eeq
For the phantom brane, this definition gives
\beq \label{effective dark fluid h}
\frac{\rho_\d}{3m^2 H_0^2} = h^2 -\frac{\rho_m}{3m^2 H_0^2} = \Omega_\sigma - 2\sqrt{\Omega_\ell}\,h
\eeq
and
\beq \label{effective dark fluid equation of state h}
w_\d =-1 + \frac{h^2+2\sqrt{\Omega_\ell}\, h-\Omega_\sigma}{2\sqrt{\Omega_\ell}\, h-\Omega_\sigma}\l(\frac{\sqrt{\Omega_\ell}}{\sqrt{\Omega_\ell}+h}\r)\,,
\eeq
where $w_\d=p_\d/\rho_\d$ is the equation of state (EOS) of the effective dark energy.

Note that the temporal (or redshift) evolution of $w_\d$ has a pole at the moment when $h=\dfrac{\Omega_\sigma}{2\sqrt{\Omega_\ell}}$. The corresponding values of redshift can be calculated from \eqref{background normal branch dimensionless}:
\beq\label{pole h z}
z_p= \left( \frac{\Omega^2_\sigma}{4\Omega_{m,0}\,\Omega_\ell} \right)^{1/3} -1 \,,~~\rm{when}~~ h_p=\dfrac{\Omega_\sigma}{2\sqrt{\Omega_\ell}}\,.
\eeq
In the domain $z>z_p$, we have $h>h_p$. Consequently, $\rho_\d<0$ and $w_\d>-1$ during this period of time in the past.
After crossing the pole, when $z<z_p$, we have $\rho_\d>0$. The effective equation of state in this region demonstrates the phantomlike behavior $w_\d<-1$ under the condition
 \beq\label{effective phantom condition h}
\sqrt{\Omega_\sigma+\Omega_\ell}- \sqrt{\Omega_\ell} < h < h_p\,.
 \eeq
 In particular, at the present moment of time (when $z=0$ and $h=1$) we have
\beq \label{effective dark fluid equation of state present}
w_\d(z=0) =-1 - \frac{\Omega_{m,0}}{1-\Omega_{m,0}}\l(\frac{\sqrt{\Omega_\ell}}{\sqrt{\Omega_\ell}+1}\r)<-1\,.
\eeq

The phantomlike behavior at the present epoch is the key feature of the phantom braneworld model.
Note that, in contrast to several phantom models, the phantom brane smoothly evolves to a de Sitter stage without running into a future singularity \cite{Sahni:2002dx, Lue:2004za}.


\subsection{Hubble evolution in terms of the variable $\Omega_m$}\label{sec: brane background Omega m}

\begin{figure}[htb]
	\begin{center}
	\includegraphics[width=0.65\textwidth]{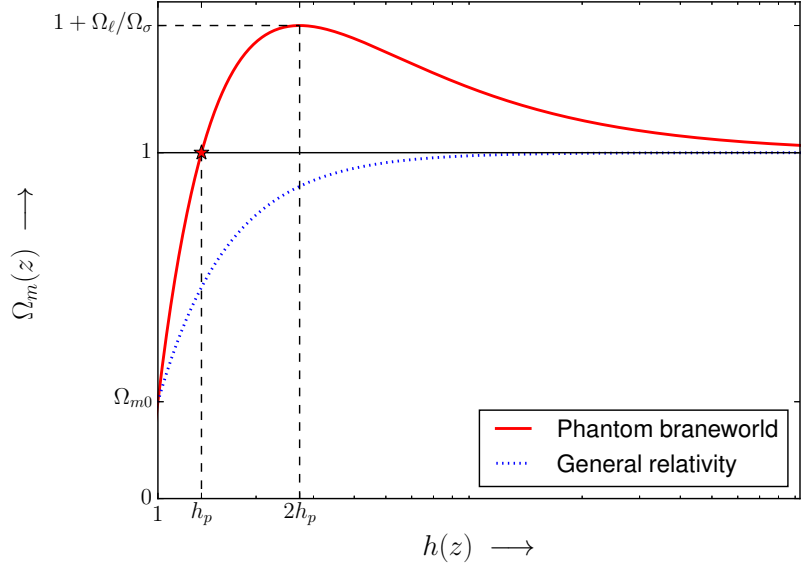}
	\end{center}
	\caption{The red curve illustrates the dependence of the variable $\Omega_m$ on the dimensionless Hubble parameter $h=H/H_0$ on the phantom brane [see relation \eqref{Omega_m brane X}]. The quantity $h$ is a monotonically growing function of redshift $z$,  so that any value of $h$ can be related to the corresponding value of $z$ via  \eqref{background normal branch dimensionless}. The red star  indicates the position of the pole in the EOS of the effective dark energy at $h=h_p=\dfrac{\Omega_\sigma}{2\sqrt{\Omega_\ell}}$ which corresponds to $\Omega_m=1$. For comparison, we have also shown $\Omega_m$ as a function of $h$ in general relativity by the blue dotted curve.}
	\label{fig: epsilon h}
\end{figure}

An interesting expression describing
\beq \label{Omega_m brane definition}
\Omega_m \equiv \frac{\rho_m}{3m^2H^2}\,.
\eeq
as a function of the expansion history, $h$, emerges from
from \eqref{background normal branch ell H}:
\beq \label{Omega_m brane X}
\Omega_m = 1 - \frac{\Omega_\sigma}{h^2} + \frac{2\sqrt{\Omega_\ell}}{h} \,.
\eeq
The relation $\Omega_m(h)$ given by this equation is illustrated in Fig.~\ref{fig: epsilon h}. Note that, in contrast to general relativity, $\Omega_m$ {\em is not a monotonic function}
 of $h$ on the phantom brane. Instead, $\Omega_m$ possesses a maximum at
$h=2h_p = {\Omega_\sigma}/{\sqrt{\Omega_\ell}}$ (crossing $\Omega_m=1$ at $h=h_p={\Omega_\sigma}/{2\sqrt{\Omega_\ell}}$). With increasing redshift, $\Om$ first increases to this maximum value
 and then decreases to unity as $h \to \infty$.

The energy density and equation of state of the effective dark energy \eqref{brane effective} in terms of the variable $\Omega_m$ can be expressed as
\begin{align} \label{effective energy Omega m}
\frac{\rho_\d}{3m^2} &= H^2(1-\Omega_m)\,, \\
\label{effective dark fluid equation of state}
w_\d &= -1 - \l(\frac{1}{1+\ell
	H}\r)\frac{\Omega_m}{1-\Omega_m}\,.
\end{align}
Note that the point $\Om(z_p)=1$ represents a pole in the EOS of the effective dark energy. Dark energy is phantomlike ($w_\d<-1$) in the region $\Omega_m(z)<1$, and quintessencelike ($w_\d>-1$)
 when $\Omega_m(z)>1$.


\section{Evolution of the growth rate}\label{sec: brane growth rate}

The theory of cosmological perturbations on the brane is quite involved because of the
presence of a large extra dimension. In particular, the bulk gravitational effects can lead to a nonlocal character of the resulting equations on the brane. Fortunately, the description of perturbations on sub-Hubble scales can be significantly simplified by using the quasistatic approximation \cite{KM} which is based on the assumption of slow temporal evolution of the five-dimensional perturbations on sub-Hubble spatial scales.
The validity of the quasistatic approximation for the phantom brane model was established in
\cite{Viznyuk:2013ywa, Bag:2016tvc}.

Evolution of the matter density contrast $\delta_m=\delta\rho_m/\rho_m$ for the braneworld model in the quasistatic approximation is given by
\beq\label{perturbation brane}
\ddot{\delta}_m+2H\dot{\delta}_m=
\frac{g_\d\rho_m\delta_m}{2m^2}\,,
\eeq
where $g_\d$ is a time-dependent function that can be regarded as a renormalization factor for the gravitational constant. For the phantom brane, it is given by the relation
\beq\label{brane grav const def}
\g_\d = 1+\frac{1}{3\mu}
\eeq
with
\beq\label{mu def}
\mu = 1+\ell H\l(1+\frac{\dot{H}}{3H^2}\r)= 1+\frac{\ell H}{2}\Bigl[1-w_\d(1-\Omega_m)\Bigr] \,,
\eeq
where $w_\d$ is the effective equation of state of dark energy, given by \eqref{effective dark fluid equation of state}.

We introduce the growth rate $f$ following \cite{Wang:1998gt}:
\beq\label{growth index definition}
f \equiv \frac{\dd \ln{\delta_m}}{\dd\ln{a}}\,.
\eeq
Its evolution can easily be determined from \eqref{perturbation brane}:
\beq\label{growth f x}
\frac{\dd f}{\dd\ln{a}} + f^2 + \l(2+\frac{\dot{H}}{H^2}\r) f = \frac{g_\d\rho_m}{2 m^2
	H^2}\,.
\eeq
We are interested in the behavior of $f$ as a function of $\Omega_m$.  In terms of the new variable $\Omega_m$, Eq. \eqref{growth f x} becomes
\beq\label{growth index brane Omega m}
6\,w_\d\,\Omega_m(1-\Omega_m) \frac{\dd f}{\dd \Omega_m}+ 2f^2 +
\left[1-3w_\d(1-\Omega_m)\right] f = 3\,g_\d \Omega_m\,,
\eeq
where we have used the relation
\beq\label{Omega_m QCDM differencial}
\frac{\dd\,\Omega_m}{\dd\ln{a}} = 3 w_\d \,\Omega_m(1-\Omega_m)\,.
\eeq

Braneworld-specific effects in \eqref{growth index brane Omega m}
are encoded in
 the effective equation of state $w_\d$. An additional modification,
specific for the braneworld model, comes from the factor $g_\d$, which renormalizes the gravitational constant for cosmological perturbations.

Note that the point $\Omega_m=1$, where the equation of state of the effective dark energy has a pole, is a regular point for the differential equation \eqref{growth index brane Omega m}. So we do not expect any singularity in the behavior of $f$ at the moment of crossing the pole.


\subsection{Series expansion in the asymptotic past}\label{sec: brane growth index series}

To find a solution of \eqref{growth index brane Omega m} in the $\Lambda$CDM model, one applies the method of series expansion around $\Omega_m=1$. In this case, the limit $\Omega_m\rightarrow1$ is equivalent to $z\rightarrow \infty$, and the corresponding series expansion gives the best result for the values $z\gg 1$. Still, parametrizing the asymptotic solution as
\beq\label{growth rate parametr GR}
f=\Omega_{m}^{\gamma} \,,
\eeq
one finds that the growth index $\gamma$ varies very slowly with $\Omega_m$.  Thus, solution \eqref{growth rate parametr GR} with $\gamma_0\approx 6/11$ describes the behavior of the growth rate with a sufficient accuracy  in the whole range $z\geqslant 0$  \cite{Wang:1998gt, Polarski:2016ieb}.

We might expect a similar result for the braneworld model. However, repeating the general-relativistic analysis is impossible in this case because $\Omega_m$ is a nonmonotonic variable in the braneworld model. It crosses the value $\Omega_m=1$ when $z=z_p$, and then tends again to $\Omega_m=1$ as $z\rightarrow \infty$ (see Sec.~\ref{sec: brane background Omega m}).

Consequently, if we wish to describe the behavior of the growth rate in the whole range $z\geqslant 0$, another variable should be chosen. Requiring this variable to be monotonic for $z\geqslant 0$, we choose it to be
\beq \label{x h definition}
\q \equiv \frac{1}{h(z)}  \,,
\eeq
where $h(z)$ is the dimensionless Hubble parameter \eqref{background normal branch dimensionless}.
The specific feature of this new variable is that $x$ is a monotonically decreasing function of redshift $z$, with $x(0) = 1$ and $x(z) \rightarrow 0$ as $z \to \infty$.

Our idea is to find the series expansion for $f(x)$ around $x=0$. We hope that, properly parametrizing the solution, we will be able to use this expansion to describe the behavior of the growth rate at the present epoch (corresponding to values of $x$ close to unity).

In terms of the variable $x$, we have
\beq \label{epsilon x}
\frac{1}{\ell H} = \sqrt{\Omega_\ell} \,x   \,,\qquad \Omega_m = 1+2\sqrt{\Omega_\ell}\, \q-\Omega_\sigma \q^2 \,.
\eeq
The evolution of the growth rate is now given by
\begin{equation}\label{growth index brane x}
3x\Omega_m(x) \frac{\dd f}{\dd x}+ 2\l(1+\sqrt{\Omega_\ell} \,x\r)f^2 +
\left(1-2\sqrt{\Omega_\ell} \,x+3\Omega_\sigma x^2\right) f 
 = 3\l(1+\sqrt{\Omega_\ell} \,x\r)\Omega_m(x)\,g_\d(x) \,,
\end{equation}
where
\beq \label{effective dark fluid equation of state x}
\g_\d = 1+\frac{2\sqrt{\Omega_\ell} \,x(1+\sqrt{\Omega_\ell} \,x)}{3\l[(1+\sqrt{\Omega_\ell} \,x)^2 + (\Omega_\sigma +\Omega_\ell)x^2\r]}\,,
\eeq
and $\Omega_m(x)$ is defined by \eqref{epsilon x}.

In the following, we suppose
that $x$ is sufficiently small\footnote{Surely, the limit $x\to 0$ ($z\to\infty$) would be rather formal here, because Eq. \eqref{growth index brane x} is valid only for matter-dominated epoch, where the quasistatic approximation can be applied.} so that both conditions
\beq \label{x lesssim 1}
\Omega_\sigma x^2 \ll 1 ~~\rm{and}~~ \sqrt{\Omega_\ell}\,x \ll 1
\eeq
are satisfied.
We seek the solution of \eqref{growth index brane x} in the form of a series expansion:
\beq\label{growth rate series}
f \sim 1 + f_1\Omega_\sigma x^2 + f_2\Omega^2_\sigma x^4 + f_3\sqrt{\Omega_\ell} \,x + f_4\Omega_\ell x^2 + f_5\Omega_\sigma\sqrt{\Omega_\ell} \,x^3 \,.
\eeq
Performing the series expansions of  \eqref{growth index brane x} to the same order, we determine the coefficients:
\beq\label{f1-f5 values}
f_1 = -\,\frac{6}{11} \,,\quad f_2 = -\,\frac{18\cdot 15}{17\cdot 11^2} \,,\quad f_3 = \,\frac{11}{8} \,,\quad f_4 = -\,\frac{153}{32\cdot 11} \,,\quad f_5 = \frac{20}{77} \,.
\eeq
This solves the problem of finding a series expansion for the growth rate in the asymptotic past, where $z \gg 1$ and $x \ll 1$. Now we study the possibility to parametrize the behavior of the growth rate in a way that extends this result to the broader range $x\leqslant 1$.


\subsection{Parametrization of the growth rate in the asymptotic past}\label{sec: parametr gamma beta}

In analogy with the $\Lambda$CDM model, we expect that the perturbative expansion of the growth rate can be represented as the power of some expression with a slowly varying exponent. We will now try to find the correct parametrization of this form that fits the series expansion obtained in the previous section.

We consider the following ansatz for the growth rate:
\beq\label{growth parametr gamma beta}
f =  \Omega_{m}^{\gamma}\Bigl(1+b\sqrt{\Omega_\ell}\,x\Bigr)^{\beta} \,,
\eeq
with
\beq\label{growth parametr gamma non const def}
\gamma = \gamma_0+\gamma_1 (1-\Omega_m)+\gamma_2\sqrt{\Omega_\ell}\,x\,, \qquad
\beta=\beta_0+\beta_1 \sqrt{\Omega_\ell}\,x \,.
\eeq
This expression is a natural generalization of the $\Lambda$CDM parametrization $f =  \Omega_{m}^{\gamma}$. The coefficient $\gamma_2$ here describes the correction of the general-relativistic growth index $\gamma$ due to brane effects. But, as we are about to show, modification of the growth index is not enough to describe the behavior of the growth rate in the braneworld model. Therefore, we introduce here the additional factor $\Bigl(1+b\sqrt{\Omega_\ell}\,x\Bigr)^{\beta}$, which becomes unity in the general-relativistic limit $\Omega_{\ell}\rightarrow 0$. The importance of this factor for the braneworld model is determined by the values of the coefficients $\beta_0$, $\beta_1$, and $b$, which will now be established.

Using \eqref{epsilon x}, we perform a series expansion of \eqref{growth parametr gamma beta}, valid under conditions \eqref{x lesssim 1}:
\beq\label{growth parametrgamma beta non const series}
f \sim 1 + \tilde{f}_1\Omega_\sigma x^2 + \tilde{f}_2\Omega^2_\sigma x^4 + \tilde{f}_3\sqrt{\Omega_\ell} \,x + \tilde{f}_4\Omega_\ell x^2 + \tilde{f}_5\Omega_\sigma\sqrt{\Omega_\ell} \,x^3 \,,
\eeq
where
\begin{align}\label{tilde f1 gamma beta non const}
\tilde{f}_1 &= -\gamma_0 \,, \\
\label{tilde f2 gamma beta non const} \tilde{f}_2 &= \,\frac{\gamma_0(\gamma_0-1)}{2}-\gamma_1 \,,\\
\label{tilde f3 gamma beta non const} \tilde{f}_3 &= 2\gamma_0+b\beta_0  \,,  \\
\label{tilde f4 gamma beta non const} \tilde{f}_4 &= 2\gamma_2-4\gamma_1+ 2\gamma_0(\gamma_0-1)+2b\beta_0\gamma_0 + b\beta_1+ \frac{b^2\beta_0(\beta_0-1)}{2} \,,\\
\label{tilde f5 gamma beta non const} \tilde{f}_5 &= -\gamma_2+4\gamma_1- 2\gamma_0(\gamma_0-1)-b\beta_0\gamma_0 \,.
\end{align}
Now we compare these coefficients with the numerical values \eqref{f1-f5 values}.
Naturally, we start with the condition that the coefficients corresponding to linear terms coincide, namely, $\tilde{f}_1=f_1$ and $\tilde{f}_3=f_3$. This results in
\beq\label{f1f3 equality no const}
\gamma_0 = \frac{6}{11}\approx 0.54545 \,, \qquad b\beta_0 = \frac{25}{88}\approx 0.28409 \,.
\eeq
Now, from the conditions $\tilde{f}_2=f_2$ and $\tilde{f}_5=f_5$, we determine
\beq\label{f2f5 equality no const}
\gamma_1 \approx 0.00729 \,, \qquad\gamma_2 \approx 0.11034\,.
\eeq
Finally, from $\tilde{f}_4=f_4$, we have
\beq\label{varphi value}
\beta_1 \approx 0.142045-\frac{0.48057}{b}\,.
\eeq

We see that $\beta_1$ can be made zero if we choose $b\approx3.383$.
In this case, we have
\beq\label{betas no const}
\beta_0 \approx 0.084\,,\qquad \beta_1\approx 0\,.
\eeq
So, finally, our parametrization is
\beq\label{growth parametr gamma beta non const fin}
f =  \Omega_{m}^{\gamma}\Bigl(1+b\sqrt{\Omega_\ell}\,x\Bigr)^{\beta} \,,
\eeq
with
\beq\label{growth parametr gamma non const fin}
\gamma \approx 0.54545+0.00729\, (1 - \Omega_m)+\alpha_0\sqrt{\Omega_\ell}\,x
\eeq
and
\beq\label{growth parametr gamma non const fin param}
\alpha_0\approx
0.11034\,,\qquad
b\approx  3.383\,, \qquad
\beta\approx  0.084 \,.
\eeq

The analytical parametrization of the growth rate given by \eqref{growth parametr gamma beta non const fin}--\eqref{growth parametr gamma non const fin param} is expected to be valid at early times, when $\Om$ is close to unity, hence $x$ is quite small. We have no reasons to believe that \eqref{growth parametr gamma beta non const fin}--\eqref{growth parametr gamma non const fin param} will be valid for $x \sim 1$. Recall that $\Omega_m$ is not a monotonic function, and this fact may result in a different parametrization for $x \sim 1$.



\begin{figure}[hbt]
\centering
\subfigure[$\oml=0.025$]{
\includegraphics[width=0.483\textwidth]{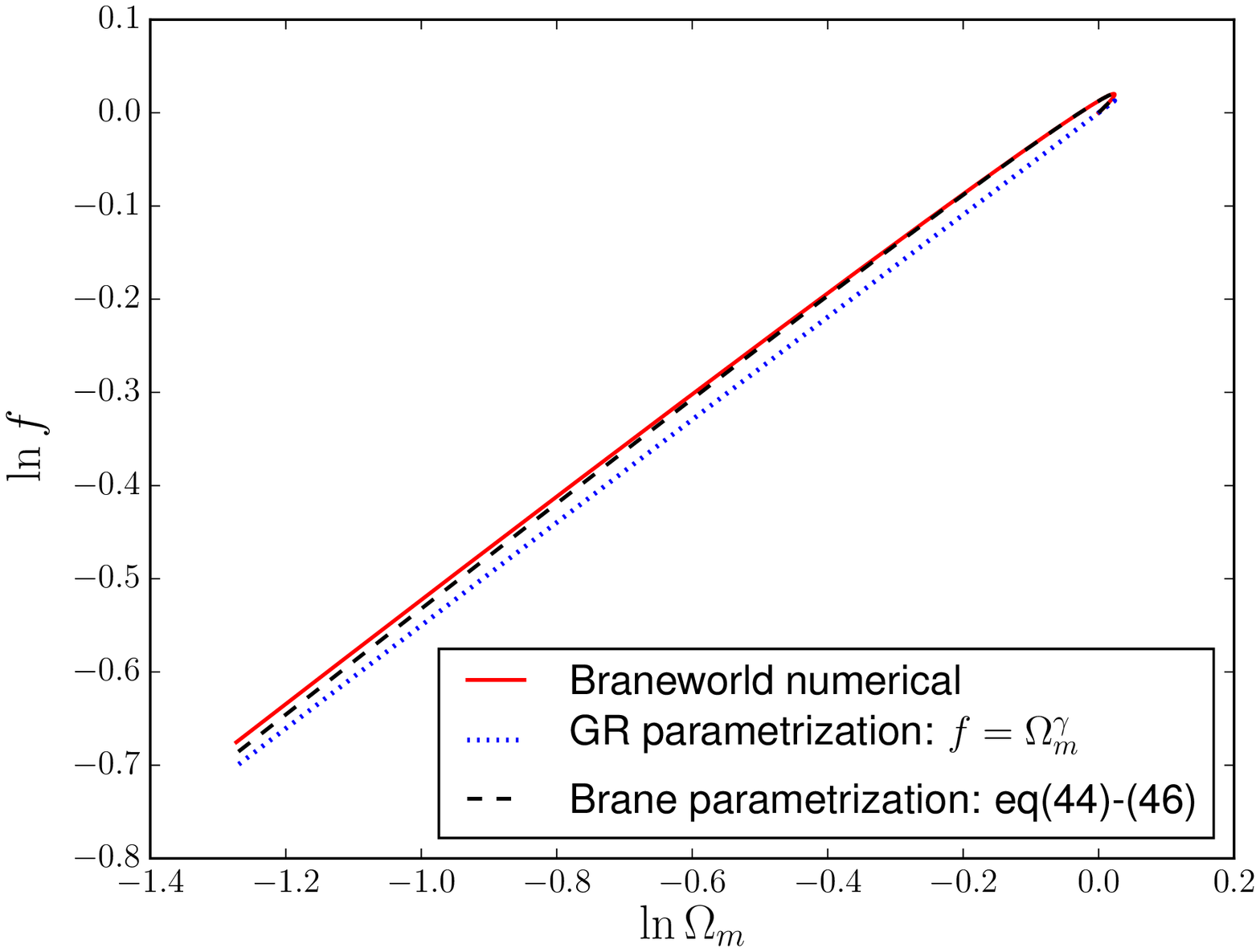}\label{fig:match:ol25}}
\subfigure[$\oml=0.2$]{
\includegraphics[width=0.483\textwidth]{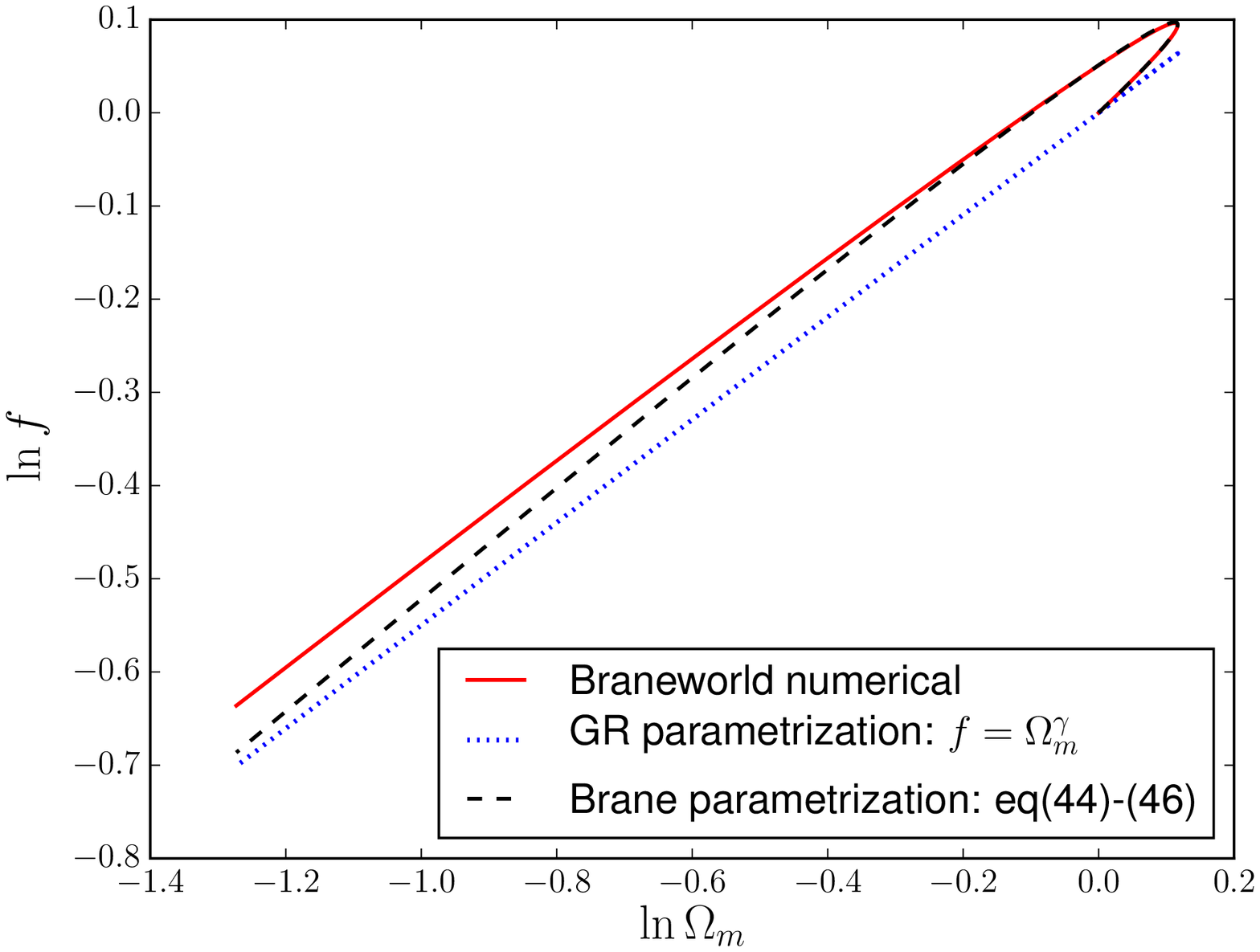}\label{fig:match:ol200}}
\caption{The analytical expression for $f$ given in \eqref{growth parametr gamma beta non const fin}--\eqref{growth parametr gamma non const fin param} (shown by the dashed black curves), is compared with the numerical solution for the growth rate (red line) for two values of the brane parameter $\oml$. For comparison, we also present here the parametrization $f=\Om^\gamma$ (blue dotted curves) with the general-relativistic parameter $\gamma$ given by \eqref{growth parametr gamma non const fin} with $\Omega_{\ell}=0$. The curves start from the matter domination, characterized by $\Om=1=f$ and end at the present epoch when $\Om=\om0$. We assume $\om0=0.28$  for illustration purposes. As is expected, the analytical solution \eqref{growth parametr gamma beta non const fin}--\eqref{growth parametr gamma non const fin param} is very accurate at early times. At late times, solution \eqref{growth parametr gamma beta non const fin}--\eqref{growth parametr gamma non const fin param} starts to deviate from the numerical solution as $\Om$ becomes significantly lower than unity. As we increase $\oml$, this deviation becomes more profound at the present epoch.
 Note that the numerical curve is multivalued for $\Om \geqslant 1$. For $\Omega_\ell=0.2$, this corresponds to $z\gtrsim 1.27$. For $\Omega_\ell=0.025$, the growth rate becomes multivalued for $z\gtrsim 2.27$.
}
\label{fig:match1}
\end{figure}

Let us compare the analytical expression \eqref{growth parametr gamma beta non const fin}--\eqref{growth parametr gamma non const fin param} with the solution for $f$ obtained by the numerical integration of \eqref{growth index brane Omega m} (see Fig.~\ref{fig:match1}). As expected, \eqref{growth parametr gamma beta non const fin}--\eqref{growth parametr gamma non const fin param} accurately matches the numerical solution at early times and becomes inaccurate at late times, when $\Om$ falls significantly below unity. We also note that, as we increase $\oml$, the analytical expression starts to deviate from the exact solution at higher values of $\Om$ (i.e., at earlier time) and the deviation at the present epoch is larger.

For comparison, we also plot in Fig.~\ref{fig:match1} the general-relativistic (GR) parametrization, $f=\Om^\gamma$.
It is evident from this illustration that the GR parametrization fails to describe the behavior of the growth rate in the braneworld model. In fact, the exact (numerical) solution for the growth rate is multivalued in the range $\Om \geqslant 1$ , $h\geqslant h_p$ (see Fig.~\ref{fig: epsilon h}). Therefore, the growth rate on the phantom brane cannot  in principle be described by the GR parametrization, $f=\Om^\gamma$, which is single-valued for all $\Om$. To correctly describe the exact solution of $f$ at all times, we need to introduce an extra factor, such as the one in \eqref{growth parametr gamma beta non const fin}.

To get closer to the parametrization valid for all times, we also need to study the future asymptote for the growth rate. This is done in the next subsection.


\subsection{Future asymptotics}\label{sec:future}
Here, we try to obtain the solution for $f$ in the distant future when  $\Om \to 0$ (or $z \to -1$). In
both GR and the phantom braneworld, $w_E \to -1$ in the distant future. Using a simple trial solution
\begin{equation}\label{eq:trail}
 f = C\, \Om^\gamma \;,
\end{equation}
one can find from \eqref{growth index brane Omega m} that $\gamma=2/3$ in both general relativity and
the braneworld model, by neglecting terms proportional to $f^2$ and $\Om$. Indeed, in the limit $\Om \to 0$, one can neglect $\Om^{2 \gamma}$ and $\Om$ with respect to $\Om^{ \gamma}$ assuming $\gamma < 1$ in advance.

\begin{figure}[hbt]
\centering
\subfigure[]{
\includegraphics[width=0.483\textwidth]{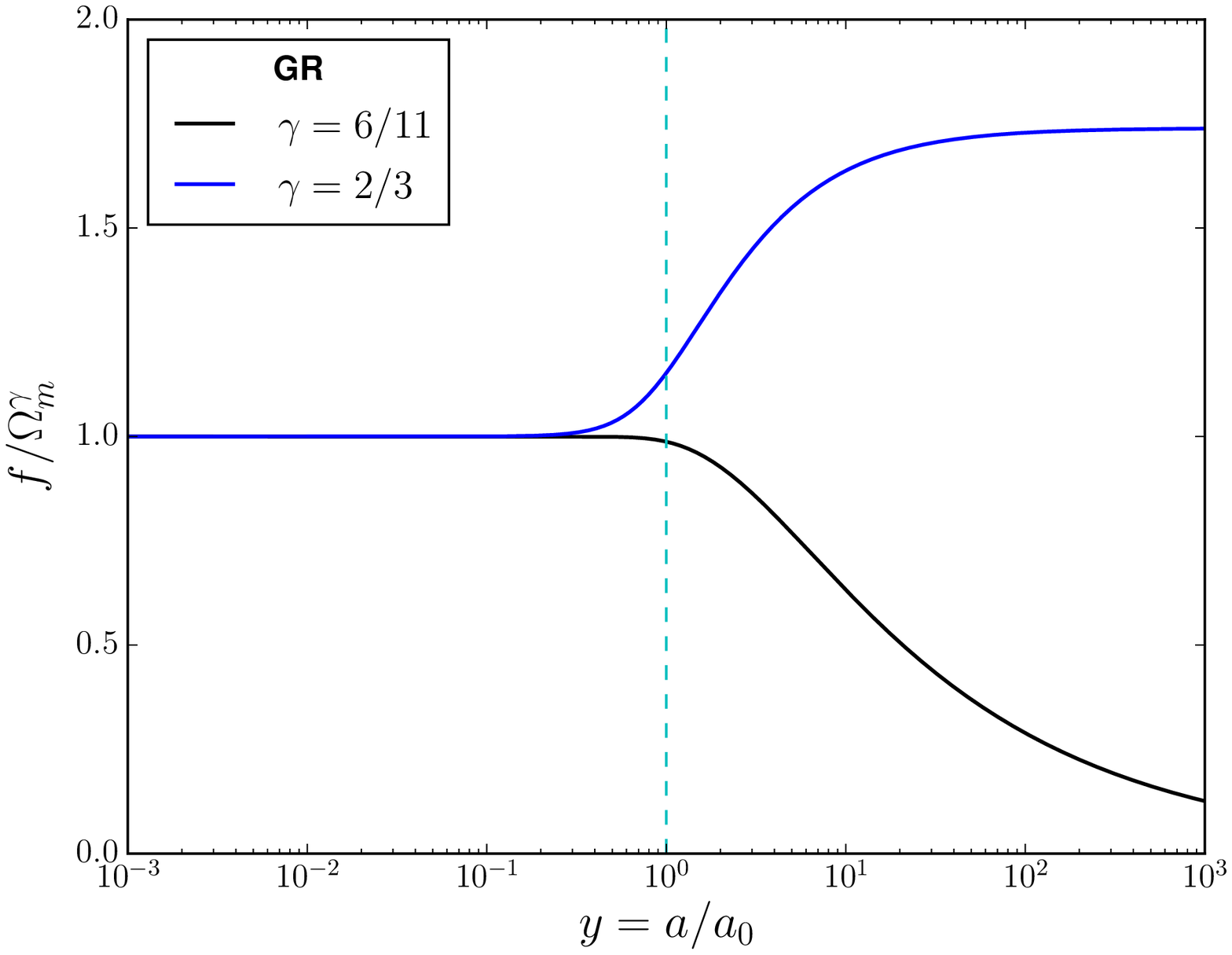}\label{fig:future_gr2}}
\subfigure[]{
\includegraphics[width=0.483\textwidth]{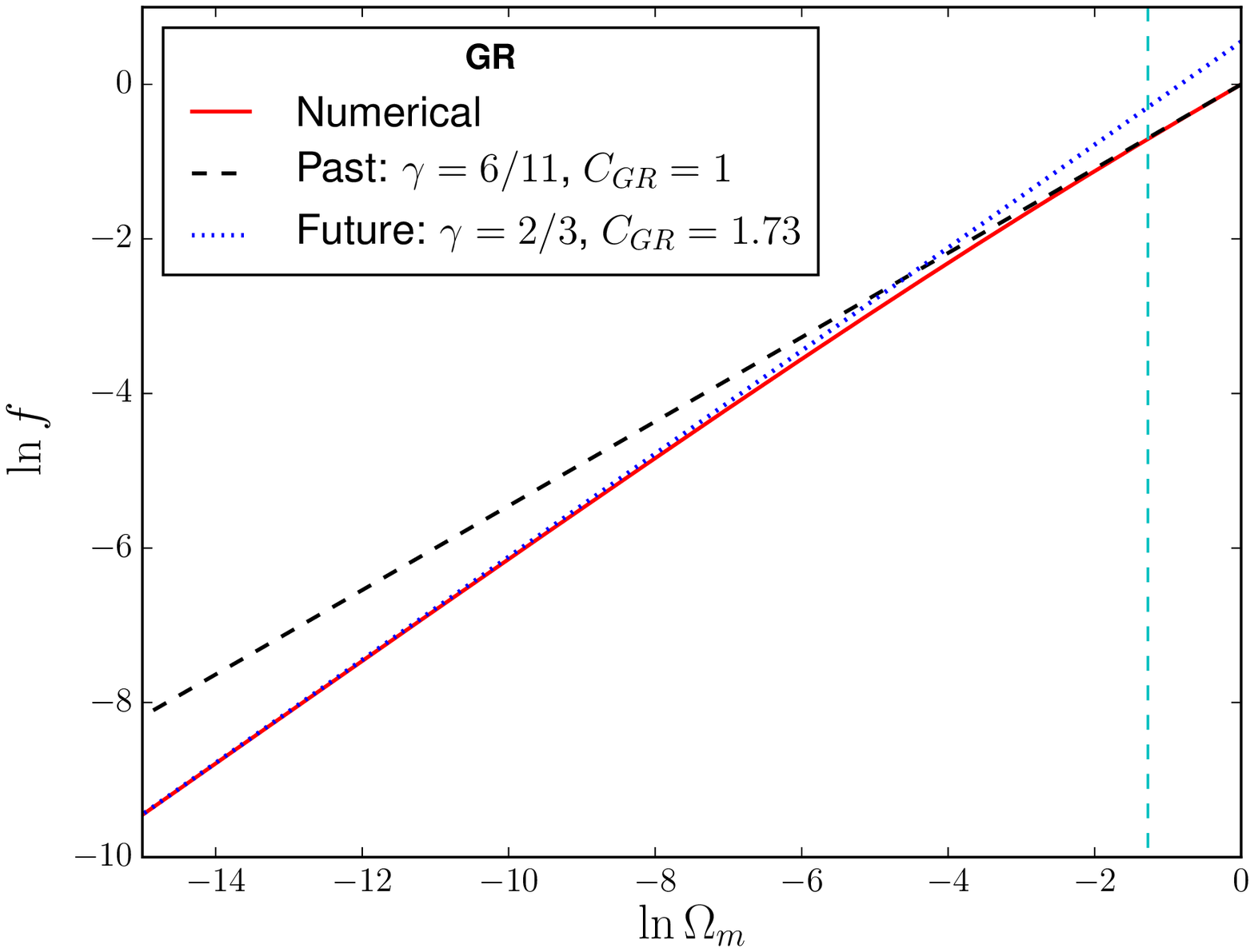}\label{fig:future_gr1}}
\caption{Both panels show that, in general relativity (GR), $f$ has two distinct solutions in the two asymptotics --- in the past and in the future. Both solutions are parametrized by $f=C\,\Om^\gamma$ with different values of the parameter $\gamma$ and the constant $C$. The dashed vertical cyan lines represent the present epoch. In the left panel, $f$ is calculated numerically and is represented by the red curve in the right panel. The blue curve in the left panel is flat in the distant future, i.e., for $a/a_0 \gg 1$. This reveals that, in the future asymptotics, $f \propto \Om^{2/3}$ where the proportionality constant is estimated as $C_{\rm GR} \approx 1.73$.
On the other hand, deep in the matter domination, $\Om \to 1$ and $f \to 1$. Therefore, both the black and blue curves in the left panel approach unity when $a/a_0 \ll 1$. In the right panel, we compare the two asymptotic solutions for $f$ with the numerical solution. The past asymptotic solution, $f \approx \Om^{6/11}$, is reasonably valid near the present epoch in GR, as is evident from both panels.}
\label{fig:future_gr}
\end{figure}

The constant $C$ in \eqref{eq:trail} can be determined from the numerical integration of \eqref{growth index brane Omega m}. In general relativity, we  get
\begin{equation}\label{eq:GR_future}
  f_{\rm GR}\approx C_{\rm GR}\, \Om^{2/3} \quad \text{in the limit} \quad \Om\rightarrow 0\,,
\end{equation}
where the constant $C_{\rm GR} \approx 1.73$ is quite robust under variation of the parameter $\om0$. The behavior of the growth rate in the asymptotic future in general relativity is illustrated in Fig.~\ref{fig:future_gr}.

For the phantom braneworld model, we found the following asymptotic solution:
\begin{equation}\label{eq:BW_future1}
f_{\rm BW}=C_{\rm BW}\, \Om^{2/3} \quad \text{in the limit} \quad \Om\rightarrow 0\,.
\end{equation}
Numerical analysis reveals that the constant $C_{\rm BW}$ here can be related to the general-relativistic value $C_{\rm GR}$ as follows:
\begin{equation}
 \frac{C_{\rm BW}}{(1+b \sqrt{\oml}x)^\beta} \approx 1.73 = C_{\rm GR}\;,
\end{equation}
which is valid for a huge range of $\oml$.
Note that $h=1/x \approx (\sqrt{\Os +\oml}-\sqrt{\oml})$ is almost constant in the limit $\Om \to 0$.

Thus one can conclude that, in the future asymptotics on the phantom brane, we have
\begin{equation}\label{eq:BW_future}
 f=C_{\rm GR}\,\Om ^\gamma (1+b \sqrt{\oml}x)^\beta \,,
\end{equation}
where $C_{\rm GR} \approx 1.73$, $\gamma\approx 2/3$, and
$\beta$ and $b$ are the same as in \eqref{growth parametr gamma non const fin param} and \eqref{eq:ansatz1}: $\beta = 0.084$, $b=3.383$.
Therefore, in both the past and future asymptotics, the growth rate on the phantom brane behaves as
\begin{equation}\label{eq:BW_GR}
 f_{\rm BW} \approx f_{\rm GR} \times (1+b \sqrt{\oml}x)^\beta \,.
\end{equation}

The deviation of the past asymptotic analytical solution for $f$, given in \eqref{growth parametr gamma beta non const fin}--\eqref{growth parametr gamma non const fin param}, from the exact solution near the present epoch can be attributed to the smooth transition from the past asymptotic solution to the future one. Even in GR, we notice that $f$ behaves entirely different in the two asymptotics --- in the past and in the future.  It is impossible to obtain a single expression for $f$ in terms of $\Omega_m$ which would be valid in the entire range of evolution, even in GR\@. Fortunately, in GR, the past asymptotic solution does not deviate much from the exact one near the present epoch, and one can compensate for this small deviation by adding higher-order correction terms to $\gamma$. Nevertheless, as we have seen in the previous subsection, adding higher-order correction terms to $\gamma$ on phantom braneworld does not significantly improve the accuracy of parametrization \eqref{growth parametr gamma beta non const fin}--\eqref{growth parametr gamma non const fin param} at late times.

Remarkably, the growth rate on the phantom brane acquires the same additional multiplicative factor $ (1+b \sqrt{\oml}x)^\beta$ to its GR counterpart in both the past and the future asymptotics. So we can expect that the parametrization
\begin{equation}\label{eq:BW_GR parametr}
 f = \Om^\gamma (1+b \sqrt{\oml}x)^\beta \,.
\end{equation}
will be reasonably valid at all times (past--present--future). This assumption will be confirmed by the numerical simulations described in the next section.


\section{Universal parametrization of the growth rate}\label{sec:numerical}

\begin{figure}[hbt]
\centering
\subfigure[$\oml=0.025$]{
\includegraphics[width=0.483\textwidth]{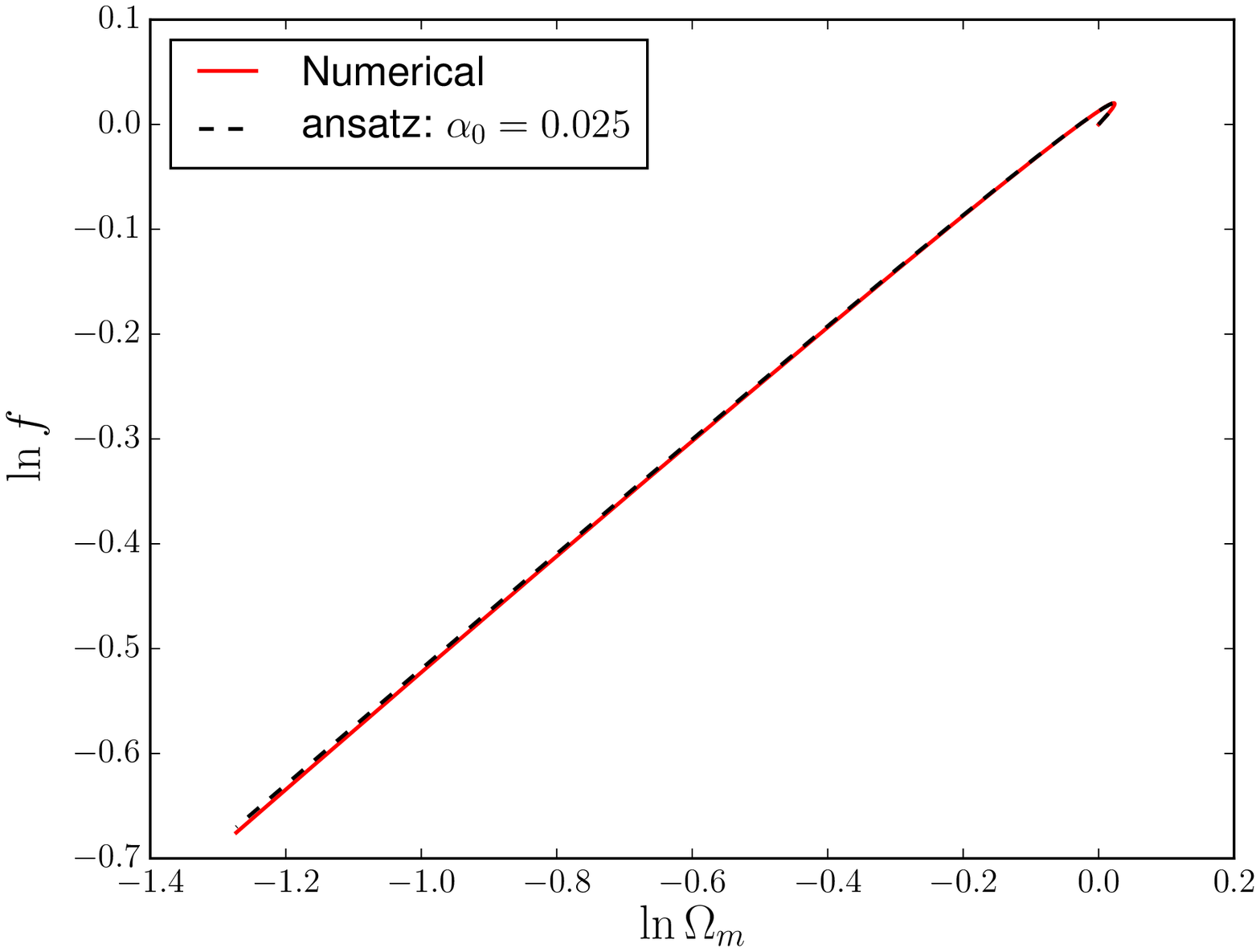}\label{fig:ansatz_ol25}}
\subfigure[$\oml=0.2$]{
\includegraphics[width=0.483\textwidth]{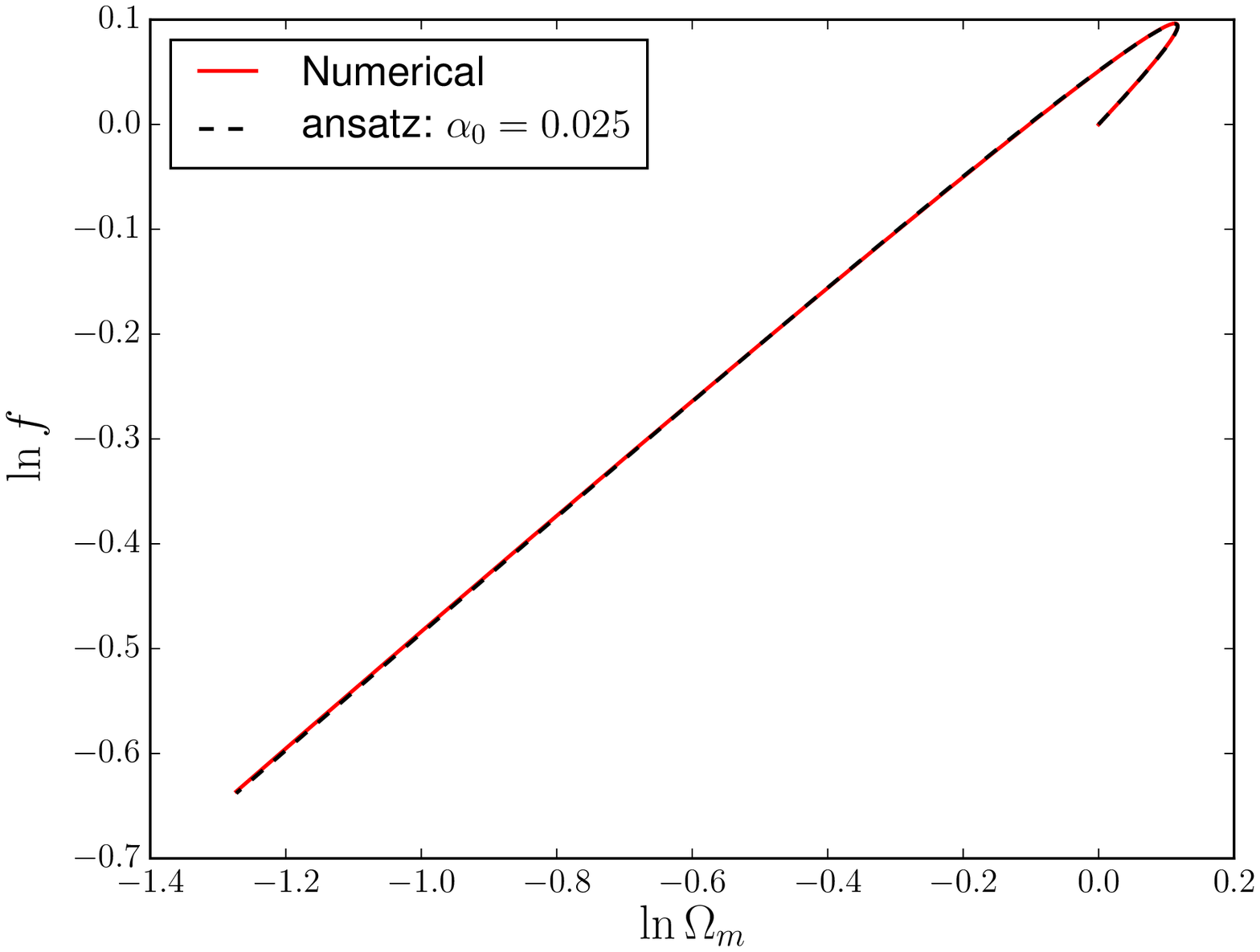}\label{fig:ansatz_ol200}}
\caption{Ansatz \eqref{eq:ansatz1}--\eqref{eq:ansatz1 gamma}  (dashed black curves) is compared with the numerical solution (red curves) for two values of the brane parameter $\oml$, starting from the matter domination until the present epoch. We find that the ansatz is quite accurate for braneworld effects as strong as $\oml=2.0$.}
\label{fig:ansatz11}
\end{figure}

As we demonstrated in the previous section, the growth rate $f$ for the phantom brane model behaves differently at different asymptotics. Parametrization \eqref{growth parametr gamma beta non const fin}--\eqref{growth parametr gamma non const fin param} works well in the asymptotic past, but fails to fit a numerical solution at the present epoch.
Therefore, in this section we seek for an ansatz that describes the evolution of $f$ reasonably well in the past, at least till the present epoch.

We have tried to get better fit with the numerical result by changing the values of the parameters in \eqref{growth parametr gamma beta non const fin}--\eqref{growth parametr gamma non const fin param}. In this way, we were able to find the ansatz that gives excellent match to the numerical solutions till the present epoch.
The ansatz is given by

\begin{figure}[hbt]
\centering
\includegraphics[width=0.65\textwidth]{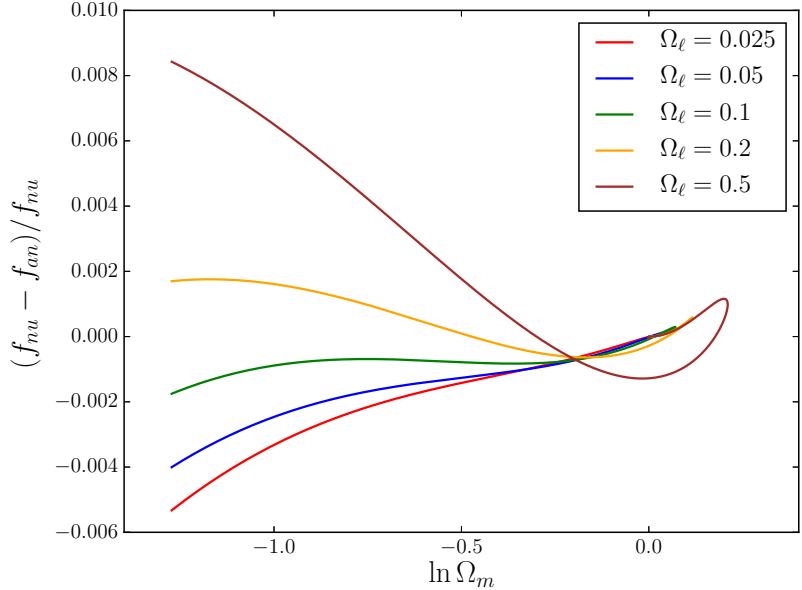}\label{fig:error_ol25}
\caption{Percentage error of the ansatz  \eqref{eq:ansatz1}--\eqref{eq:ansatz1 gamma} is shown for  different values of the brane parameter $\oml$, starting from the matter domination until the present epoch. Here, $f_{nu}$ and $f_{an}$ represent the numerical solution for $f$ and the ansatz \eqref{eq:ansatz1} respectively. Note that, during matter domination $\Om \to 1$ and, at the present epoch, $\Om=\om0=0.28$. }
\label{fig:error1}
\end{figure}

\begin{equation}\label{eq:ansatz1}
 f =  \Omega_{m}^{\gamma}\Bigl(1+b\sqrt{\Omega_\ell}\,x\Bigr)^{\beta} =  \Omega_m^\gamma\left(1+\frac{b}{\ell H}\right)^\beta\;,
\end{equation}
with
\begin{equation}\label{eq:ansatz1 gamma}
\gamma \approx 6/11+0.00729\, (1-\Omega_m)+\frac{\alpha_0}{\ell H} \;, \quad \text{where} \quad \alpha_0=0.025\;,
\end{equation}
and $\beta$, $b$ are same as in \eqref{growth parametr gamma non const fin param}, i.e., $\beta=0.084$, $b=3.383$.

 Ansatz \eqref{eq:ansatz1}--\eqref{eq:ansatz1 gamma} differs from \eqref{growth parametr gamma beta non const fin}--\eqref{growth parametr gamma non const fin param}, obtained by considering the past asymptotics,  by the value of a single parameter $\alpha_0$ (recall that $\alpha_0\approx 0.11$ in the exact asymptotic solution for $x \equiv 1/h \ll 1$).
 Figure \ref{fig:ansatz11} shows the comparison of ansatz \eqref{eq:ansatz1}--\eqref{eq:ansatz1 gamma}  with the exact solution.
We see that the exact solution of $f$ is reasonably well described by the ansatz at least till the present epoch $z\gtrsim 0$, even for braneworld effects as strong as $\oml=0.2$.
The error of this ansatz is of the order $0.1 \%$ for values of $\oml$ consistent with the observations. The evolution of the error for different values of $\oml$ is shown in Fig.~\ref{fig:error1}. As we can see, ansatz \eqref{eq:ansatz1}--\eqref{eq:ansatz1 gamma}  is reasonably valid till the present epoch for a wide range of $\oml$. For example, the maximum error is below  $1\%$ even for the brane parameter as large as $\oml = 0.5$.

 For comparison, we can estimate the maximal error for the standard general-relativistic parametrization $f=\Omega_m^\gamma$ from Fig.~\ref{fig:match1}. We note that the parametrization $f=\Omega_m^\gamma$ results in $f=1$ whenever $\Om=1$ for any $\gamma$. On the other hand, the exact solution for $f$ on the phantom brane reveals that $f>1$ at $z=z_p$ (the pole in the EOS) when $\Om=1$. The discrepancy of the parametrization  $f=\Omega_m^\gamma$ with the numerical solution at this point is about  $0.052$ for $\Omega_{\ell}=0.2$, and  $0.013$ for $\Omega_{\ell}=0.025$. Consequently, the maximal error of the parametrization $f=\Omega_m^\gamma$ {\em with any $\gamma (z)$\/} cannot be smaller than $5.2\%$ for $\Omega_{\ell}=0.2$, and $1.3\%$ for $\Omega_{\ell}=0.025$. For instance, the maximal errors of the parametrization with the general-relativistic $\gamma (z)$ in these cases constitute $6.4\%$ and $2.7\%$, respectively.


\section{Conclusions}\label{sec: conclusion}

Our analysis demonstrates that the growth rate for the phantom brane model can be parametrized as
 \beq\label{growth parametr gamma beta non const fin1}
f=\Omega_m^\gamma\left(1+\frac{b}{\ell H}\right)^\beta \,,
\eeq
with the growth index $\gamma$ and other parameters given in \eqref{eq:ansatz1}--\eqref{eq:ansatz1 gamma}.
 Such a form of the growth rate provides an excellent fit to numerical simulations from
very large $z$ to the present epoch. We have established that the above parametrization is highly accurate (the maximum error is of the order $0.1\%$) for a wide range of the brane parameter $\oml$.

The standard general-relativistic parametrization $f=\Om^\gamma$ fails to describe the behavior of the growth rate on the phantom brane because the exact solution for the growth rate
in this case is multivalued in the domain $\Om \geqslant 1$, whereas the function $f=\Om^\gamma$,  is single-valued for all $\Om$,  and moderate dependence of $\gamma$ on redshift cannot remedy the situation. 	 For example, the exact solution is multivalued for $z\gtrsim 1.27$ if  $\Omega_\ell=0.2$, or for $z\gtrsim 2.27$ if $\Omega_\ell=0.025$. The maximal error of the parametrization $f=\Om^\gamma$  with any $\gamma (z)$ cannot be smaller than $5.2\%$ for $\Omega_{\ell}=0.2$, and $1.3\%$ for $\Omega_{\ell}=0.025$. In the case of general-relativistic $\gamma (z)$, these errors are as large as $6.4\%$ and $2.7\%$, respectively.

A similar situation can be expected for other models of modified gravity.
Our results therefore suggest that a general parametrization of the form\footnote{To the best of
 our knowledge,  a parametrization of the form \eqref{growth parametr modified} for modified gravity models was first proposed in \cite{Resco:2017jky}.}
 \beq\label{growth parametr modified}
f=K(a)\Omega_m^\gamma(a)
\eeq
may be more suitable for the description of perturbations in
 modified gravity models than the usual approach based on $f=\Omega_m^\gamma(a)$.

\section*{Acknowledgments}

The work of A.~V. and Y.~S. was partially supported by The National Academy of
Sciences of Ukraine (project No.~0116U003191) and by grant 6F of the Department of Targeted Training of the Taras Shevchenko National University of Kiev under the National Academy of Sciences of Ukraine. S.~B. thanks the Council of Scientific and Industrial Research (CSIR), India, for
financial support as senior research fellow.  Y.~S. is grateful to the Indian National Science Academy for the award of the Professor DS Kothari Chair and acknowledges INSA and IUCAA
hospitality during his visit to IUCAA under this programme.


\end{document}